\documentclass{article}
\usepackage{geometry}                
\usepackage{graphicx}
\usepackage{amssymb}
\usepackage{epstopdf}
\usepackage{slashed}
\usepackage{hyperref}

\usepackage[customcolors]{hf-tikz}
\hfsetfillcolor{gray!15}
\hfsetbordercolor{white} 

\usepackage{url}

\usepackage{natbib}
\bibpunct{(}{)}{;}{a}{}{,}

\usepackage{footnote}
\makesavenoteenv{eqnarray}

\usepackage{tocloft}
\pdfoutput=1

\begin{document}

\title{Chapter 1: Introduction: The emergence of spacetime}
\author{Nick Huggett and Christian W\"uthrich\thanks{This is a chapter of the planned monograph \emph{Out of Nowhere: The Emergence of Spacetime in Quantum Theories of Gravity}, co-authored by Nick Huggett and Christian W\"uthrich and under contract with Oxford University Press. More information at \url{www.beyondspacetime.net}. This work was supported financially by the ACLS and the John Templeton Foundation (the views expressed are those of the authors not necessarily those of the sponsors.}}
\date{18 January 2021}                                           
\maketitle

\label{ch:intro}

\tableofcontents
\

\begin{center}
\Large{\textbf{``Big Bang Machine Could Destroy Earth"}}
\end{center}

\ 

\noindent\dots ran an attention grabbing headline in  \textit{The Sunday Times} \citep{Leake:1999fk}, regarding the new Relativistic Heavy Ion Collider (RHIC) at Brookhaven National Laboratory. To be fair, the main apocalyptic concern of the paper was that the RHIC would create a form of matter in which strange quarks eat the up and down quarks found in ordinary matter. But it also discussed the possibility that experiments involving high-energy collision of gold ions could create microscopic blackholes, which would pull all the matter in the world into them. Such scenarios were taken seriously enough that they were evaluated by a panel of elders, who concluded that the chances of any such events were utterly minuscule \citep{W.-Busza:1999uq}---happily, up to the time of writing, they have not been contradicted by events at Brookhaven! 

Let's look into the business of black hole formation more carefully to explain why physics needs an account of quantum gravity. First, the RHIC was built to probe how matter behaves under intense temperatures and pressures---in effect recreating in a tiny region the state of the universe within the first second of its existence, when quarks and gluons flowed in a plasma rather than binding to form particles. The predictions tested here are largely those of quantum chromodynamics, the quantum theory of the strong force binding nucleons and their constituents. That is, the collisions between heavy nuclei such as gold in the RHIC are governed by the laws of quantum mechanics (QM). 

The concern over black holes, however, arises when one asks how  general relativity (GR)---the classical, non-quantum, theory of gravity, gets into the picture. According to GR, the spacetime metric outside a sphere of mass $M$ takes the form: 

\begin{equation}
\mathrm{d}s^2 = c^2\mathrm{d}t^2(1-\frac{2GM}{rc^2}) - \mathrm{d}r^2/(1-\frac{2GM}{rc^2}),
\end{equation}
where $G$ is Newton's gravitational constant, $c$ is the speed of light, and $r$ the distance from the center. ($\mathrm{d}s^2$ is the infinitesimal spacetime `distance' squared---the `interval'---between two radial points separated in time by $\mathrm{d}t$ and space by $\mathrm{d}r$, in suitable co-ordinates; though an exact understanding is not crucial here.) What is important is that this quantity blows up when $2GM=rc^2$, or $r=2GM/c^2$. Understanding this occurrence was an important issue in the early development of GR, but what it actually signifies is the presence of an `event horizon' around the mass, from which neither matter nor light can escape---the boundary of a black hole. Of course this story only makes sense if the mass is all located within a radius of $2GM/c^2$, since the metric formula only holds outside the mass. So one can equally well say that if one has a mass $M$ it will only form a black hole if it is all located within a radius less than $2GM/c^2$. 

So finally, the question posed by the panel at Brookhaven was how small a region would the amount of energy to be created in collisions be located (see page 7 of the report)? Acting cautiously, they assumed the best conditions for black hole formation, supposing that all the energy produced by the collision contributes to the mass: about 50 times that of a gold atom. For a black hole of this mass the event horizon has a radius of $10^{-39}$m. On the other hand, a gold atom has a radius of around $10^{-12}$m, so even supposing that all the energy is concentrated in a region the size of a suitably Lorentz-contracted nucleus, general relativity predicts that collisions will be many, many orders of magnitude from creating a black hole.

Phew.\footnote{\label{rhicbh}Don't be confused if you have read of black holes being created at RHIC. In fact what has (perhaps) been observed is the Unruh effect, which is formally equivalent to Hawking radiation, but does not involve black holes, but acceleration, which is in a sense indistinguishable from a gravitational force according to GR. See \cite{Nas:05}.}

\section{Quantum gravity and philosophy}

What we have then is an argument that the physics of Brookhaven lies within the domain of relativistic QM, but that the gravitational effects of the collisions are utterly negligible. Perhaps the world is just `dappled' in this way: in some domains, such as the motions of the planets (GR explains the perihelion of Mercury, for instance) GR holds and QM is irrelevant; in others, as in RHIC, it is QM that holds sway, with GR  entering only to provide a background geometry determined by ambient bodies, not by the system under consideration. But the very argument here shows that it is possible to bring considerations from both theories to bear on a single system, making clear that one can sensibly ask whether there are domains in which both theories apply. On our current understanding of the universe  indeed there are. In the first place, there is the big bang, which is entailed by GR given the current state of the universe, in which matter becomes so hot and compact that QM effects will necessarily occur. (And conversely, if the inflationary hypothesis is correct, between the first $10^{-33}$ to $10^{-32}$s quantum fields provide the energy which drives the expansion of the universe, according to the laws of GR.) In the second, Hawking radiation is predicted to occur around black holes  as a QM effect resulting from the geometry of spacetime given by GR \citep{Hawking:1974fk}. Indeed, assuming that the local equivalence of acceleration and gravity---the `equivalence principle'---holds in the quantum domain, then RHIC does provide tests for this physics, since the radiation produced by decelerating ions can be measured (see footnote \ref{rhicbh}): under the assumption, it provides indirect tests of the overlap of gravity and QM.

Thus, we say, the ultimate need for a theory that in some way unifies QM and GR---a theory of \emph{quantum gravity} (QG)--- arises from the existence of phenomena in our universe in which the domains of the two overlap. There are other, more theoretical, arguments for such a theory and concerning the form it should take. For a critical evaluation of these, arguing that overlap is best reason to seek quantum gravity, and that empirical considerations best dictate its form see \cite{Callender:2001uq}; for further discussion \cite{wut05}.\footnote{This situation may change if the technology improves enough over the next few years to carry out the experiment proposed by \cite{BosMazMor:17}.}

There are then good reasons for physicists to investigate QG. This book is predicated on the view that it also has an important call on the efforts of philosophers. Indeed, we would like this work to encourage, by example, our colleagues to be more adventurous in their choice of topics of enquiry. Philosophy of physics, we suggest, has a tendency to look too much to the past, and to the metaphysics of well-established physics (and of course to internecine disputes): classical statistical mechanics, classical spacetime theory and non-relativistic quantum mechanics are `so twentieth (or even nineteenth) century', and yet have a virtual lock on the discipline. While quantum field theory (QFT) is becoming a significant topic, that is still at least half a century behind the physics!

We're overstating things somewhat for effect here: of course, even old theories do face important foundational problems, and their consequences for our broader understanding of the world take considerable elucidation. And of course it is unfair to suggest that no philosophers of physics show an interest in contemporary physics. Indeed, since we started writing this book, there has been an explosion of interest in QG, especially amongst a younger generation of scholars, which we find very exciting. Still, we do say that collectively the discipline pays insufficient attention to cutting edge physics, and hope this book in some way serves as an impetus to greater engagement.

Our point is not that novelty is good for its own sake, nor that philosophy is a `hand-maiden', who should dutifully follow the fashions of physics. Rather, we are inspired by recent work in the history and philosophy of science to believe that it is central to the business of philosophy to engage with developing physical theories---both because the search for philosophical knowledge must be responsive to empirical discoveries, and because philosophy has important contributions to make to the development of physics (and other sciences). We will discuss this point at greater length below (\S\ref{sec:rolepp}), to explain our aims and motivations. 

First, in part to make that discussion more concrete, in \S\ref{sec:WwoST} we will very briefly introduce some theories of QG (or in its vicinity). We especially want to focus on a rather generic feature of them---that in various ways they do not contain familiar spacetime at a fundamental level, but rather it `emerges' (in a sense to be discussed) in a higher, non-fundamental domain. That will lead us to a discussion (in \S\ref{sec:MsCh}) of the challenges to the very idea that something a seemingly fundamental as spacetime could be derivative, from even more fundamental, yet non-spatiotemporal, physics. We then address these challenges, analyzing how they can be overcome, and illustrating this process in a historical case (\S\ref{sec:introPS}); and in a short contemporary example (\S\ref{sec:introNCG}). This discussion encapsulates much of the work of the book: the chapters introduce, for philosophers, several proposals for a theory of QG, discuss the ways in which they eliminate spatiotemporal structures, and investigates the ways in which they are recovered as effective, apparent structures. What we propose is a form of `functionalism', so in \S\ref{sec:funcES} we explain that position.

In a final section (\S\ref{sec:finalintro}) we will give an overview of the different strategies one might take towards quantizing gravity, to relate the different proposals that we will consider in the book.

\section{Worlds without spacetime?}
\label{sec:WwoST}

All theories of QG are, to a large extent, speculative; some of the examples that follow are more speculative than others. However, as we shall explain below, there are good reasons to think that they may still teach important lessons in the search for QG. The first three examples are the focus of the following chapters; the remaining two are also illuminating, but we have discussed them elsewhere.

\begin{itemize}

  \item \textbf{Causal Set Theory (CST):} As we will see in chapters 2 and 3, CST makes liberal use of GR as a vantage point for its research programme. In fact, it takes its most important motivation from theorems stating that given the causal structure of a spacetime, its metric is determined up to a conformal factor. In other words, the causal structure determines the geometry of a spacetime---but not its `size'. Taking this cue, CST posits that the fundamental structure is a set of elementary events which are locally finite, partially ordered by a basic causal relation. In other words, the fundamental structure is a {\em causal set}. The assumption of local finitarity is nothing but the formal demand that the fundamental structure---whatever else it is---is discrete. Together with the demand of Lorenz invariance at the derived level, the discreteness of causal sets forces a rather odd locality structure onto the elementary events of the causal set (\S3.5). Furthermore, although the fundamental relation of causal precedence can double up as something akin to temporal precedence, space is altogether lost in a causal set (\S2.3). Jointly, these facts entail that the structure we are facing in causal set theory is also rather different from the spacetime encountered in GR. In fact, the quantum nature of the causal sets yet to be incorporated into causal set theory is bound to further complicate the picture and to remove the resulting structure from that of relativistic spacetimes.\\
  
    \item \textbf{Loop Quantum Gravity (LQG):} LQG starts out from a Hamiltonian formulation of GR and attempts to use a recipe for cooking up a quantum from a classical theory that has been utilized with great success in other areas of physics. This recipe is the so-called {\em canonical quantization}. The goal of applying the canonical quantization procedure is to find the physical Hilbert space, i.e.\ the space of admissible physical states, and the operators defined on it that correspond to genuinely physical quantities. As will be seen in chapters 4-5, following this recipe leads rather straightforwardly into a morass of deep conceptual, interpretative, and technical issues concerning the dynamics of the theory as well as on time quite generally. We find that the states in the `kinematic' Hilbert space afford a natural geometric interpretation: its elements are states that give rise to physical space, yet are discrete structures with a disordered locality structure. At least in one basis of this Hilbert space, the states appear to be states of a granular structure, welding together tiny `atoms' of space(time). It is crucial to this picture that these atoms of space(time) are atoms in the original meaning of the word: they are the truly indivisible smallest pieces of space(time). The smooth space(time) of the classical can thus be seen to be supplanted by a discrete quantum structure. Moreover, since generically a state of this structure will be a superposition of basis states with a determinate geometry, generic states will not possess determinate geometric properties. If continuity, locality, or determinate geometry was an essential property of spacetime, then whatever the fundamental structure is, it is not spacetime. In this sense, spacetime is eliminated from the fundamental theory.\\

  \item \textbf{String Theory:} According to string theory (chapter 6) tiny one-dimensional objects move around space, wiggling as they go---the different kinds of vibration correspond to different masses (charges and spins) and hence to different subatomic particles. So it sounds as if spacetime is built into the theory in a pretty straight-forward way; however, we will see that things are not so simple. First, in chapter 7 we will see that various versions which are intuitively very different in fact correspond to the same physics---are `\emph{dual}'. For instance, suppose that at least one of the dimensions of space is `compactified', or circular. Then it turns out that a theory in which the circumference of the dimension is $C$ has the same collection of values for physical quantities as a theory in which the circumference is $1/C$: i.e., that theories in which compactified dimensions are small are physically indistinguishable from---or `dual' to--- those in which they are large. Other dualities relate spaces of different topologies. These facts raise important questions about whether the space in which the string lives is the one we observe, since that is definitely large though the string space could be small, or even conventional. Second, in chapters 8-9, we will see how the geometrical structure of spacetime---and indeed GR---arises from the behavior of large collections of strings in a `graviton' state, the quantum particle that mediates gravitational forces. \\
  
\item \textbf{Group Field Theory:} Consider rotations, by and angle $\theta$, in the plane: a different one for each value of $0\leq \theta < 2\pi$. These form a group under composition: a rotation by $\alpha$ followed by an angle $\beta$ is just a rotation by $\alpha+\beta$ (and for example, a rotation by $2\pi-\alpha$ undoes a rotation by $\alpha$). Similarly for rotations in three dimensions, and indeed similarly for Lorentz transformations (which are in fact nothing rotations in Minkowski spacetime), and so on. For each, composition yields a different function from any pair to third. We can thus characterize the abstract group structure simply by the action of this function of an entirely arbitrary set of elements---forget that we started with rotations, and let the elements be anything, with a composition rule isomorphic to that of the rotations. One could take then a set of such blank elements, which compose like rotations in the plane, but which should not be thought of as literal rotations: no plane at all is postulated, the only manifold is that formed by the group elements themselves---the circle of points with labels $0-2\pi$, not a 2-dimensional plane.\footnote{More accurately, the space that is assumed is not the space of relativistic physics, or that of planar rotations, but four copies of the group of Minkowski rotations.} And finally one can introduce a field on this group manifold, a real number for each group element, with a dynamical law for its evolution; and indeed quantize this field. As we have emphasized, while physical space was used to guide the construction of this theory, it is not an explicit component of it, and yet models of general relativity can be derived from it. This example is discussed in greater detail in \cite{Ori:14} and \cite{Hug:18}.\\

  \item \textbf{Non-Commutative Geometry:} Picture a Euclidean rectangle whose sides lie along two coordinate axes, so that the lengths of its sides are $x$ and $y$---its area is $x\cdot y= y\cdot x$. But what if such products fail to commute, $xy\neq yx$, so that area is no longer a sensible quantity? How can we understand such a thing---a \emph{non-commutative geometry}? By abandoning ordinary images of geometry in terms of a literal space (such as the plane) and presenting it in an alternative, algebraic way. In fact, our example already starts to do so: even thinking about areas as products of co-ordinates uses Descartes' algebraic approach to Euclidean geometry. Once we have entered the realm of algebra, all kinds of possible modifications arise. Especially, an abstract algebra $\mathcal{A}$ requires an operation of `multiplication', $\star$, but this can be a quite general map from pairs of elements, $\star:\mathcal{A}\times\mathcal{A}\to\mathcal{A}$, \emph{which need not be commutative}! For instance, one could define `multiplication' to satisfy $x\star y - y\star x = \theta$ (a small number), and use it to generate polynomials in $x$ and $y$; these carry geometric information. Such a thing is perfectly comprehensible from the abstract point of view of algebra, but it cannot be given a familiar Euclidean interpretation via Cartesian geometry. So, such a theory seems to describe a world that is fundamentally algebraic, not spatial (in the ordinary sense)---there is $x\star y$ and $y\star x$ but no literal rectangle. If there is thus fundamentally nothing `in' space, is the ultimate ontology `structural', based on algebraic relations only? And how could an appearance of familiar (commutative!) space arise; especially, what significance could point-valued quantities have? We will return to this example in \S\ref{sec:MsCh}.\\

\end{itemize}

All of these examples are speculative to some extent or other, and none can claim to be a complete quantum theory of gravity (and none has convincing, currently testable, novel predictions!), yet all have some claim to model relevant physical features of QG, worth exploring. In particular, we have emphasized in each case how spacetime features are missing in the theories (to be spelled out in detail in later chapters). This situation thus appears to be a common condition of many approaches to QG, in which case we say that classical, relativistic spacetime is `emergent'. We emphasize (as we have elsewhere) that we do not use this term in its strongest philosophical sense to indicate the \emph{inexplicability} of X from Y: as some have claimed life or mind emerges from matter. On the contrary, we argue that classical spacetime structures \emph{can} be explained in more fundamental terms: indeed, it was largely to explicate how physicists do so that we wrote the book. Some might then say that spacetime `reduces' to non-spatiotemporal QG, but we prefer to stick with the notions of `explanation' or `derivation', because there are many notions of `reduction', some of which are too strict. But we are also happy to speak of (weak) `emergence' even when spacetime is derived, because the gulf between a theory that does not assume spacetime, and one that does is so great. Having spacetime or not makes a huge formal and conceptual difference, in particular because in almost all theories prior to QG classical spacetime has apparently been one of the most basic posits. Indeed, this very gulf makes one wonder what it could mean to derive spacetime, and whether it is possible at all.

Before we proceed, we need to introduce some terminology to keep the discussion straight. The issue is that the theories of QG often contain some object referred to as `space', even when they do not assume `space' in the ordinary sense. For instance, there may be a `Hilbert space', or a `dual space', or `Weyl space', or `group space'. So we will refer to spacetime in the ordinary sense as `classical', or `relativistic', or sometimes just `space' or `spacetime' when the context makes matters clear. (We eschew the phrase `physical space', since the other `spaces' may well be part of the fundamental physical furniture. We have previously used `phenomenal space', to indicate that classical space is that of observable phenomena, according to the physicist's use of `phenomenological'. However, this leads to confusion with the philosophical doctrine of `phenomenalism', so we have dropped it.)

By classical or relativistic spacetime, we mean that theorized in QM (especially QFT) and relativity, approximated in non-relativistic mechanics, and ultimately implicated in our observations of the physical world. As stated, that is not an entirely homogeneous concept, so we will say more later (\S\ref{sec:funcES}) about exactly what features of classical spacetime are emergent from our theories of QG. First, we turn to some challenges to the project of deriving spacetime.

\section{Challenges to spacetime emergence}
\label{sec:MsCh}

Space and time are so basic to both our manifest and scientific images of the world that at first the mind boggles at the thought that they might be mere `appearances' or `phenomena', of some deeper, more fundamental, non-spatiotemporal reality. Is a physics without spacetime even intelligible? And if it is, is spacetime the kind of thing whose existence could be explained? At its core, this book seeks to address these questions: on the one hand explicating the worlds described by theories of QG, while on the other showing how spacetime can be derived from them. But to understand the nature and methodology of that project, it is important here to unpack the mind boggling, vertiginous panic about the very idea. Larry \citet{skl83} gave expression to this all too common sentiment among philosophers (and physicists) when he wrote\footnote{A note on terminology: we take `Platonists' to be committed to the existence of abstract entities, such as propositions, sets, love, and justice, but also to the existence of the concrete, physical, and spatiotemporal world. Those who maintain that our world is fundamentally mathematical in nature and thus entirely consists in ultimately abstract entities or structures are often labelled as `Pythagoreans'. Since we are interested not in whether there exist abstracta, but in the possibility that all physical existence is grounded in non-spatiotemporal structures, we will refer to those who maintain that a fundamentally non-spatiotemporal physical world is not devoid of ``real being''---no doubt historically inaccurately---as \emph{Pythagoreans}. We take this Pythagoreanism to be Sklar's target---and the one of this monograph.}
\begin{quote}
What could possibly constitute a more essential, a more ineliminable, component of our conceptual framework than that ordering of phenomena which places them in space and time? The spatiality and temporality of things is, we feel, the very condition of their existing at all and having other, less primordial, features. A world devoid of color, smell or taste  we could, perhaps, imagine. Similarly a world stripped of what we take to be essential theoretical properties also seems conceivable to us. We could imagine a world without electrical charge, without the atomic constitution of matter, perhaps without matter at all. But a world not in time? A world not spatial? Except to some Platonists, I suppose, such a world seems devoid of real being altogether. (45)
\end{quote} 

According to Sklar, a non-spatiotemporal world is inconceivable, and thus presumably not even metaphysically possible, let alone physically. This monograph is concerned with establishing the possibility of a fundamentally non-spatiotemporal world, articulating the consequences of such a possibility, and defending the idea that spacetime may be merely emergent in a perfectly acceptable scientific explanation of the manifest world. So in this section we will discuss various more precise ways that one might doubt the possibility of deriving spacetime.

First, in \cite{HugWut:13} we discussed the idea that a theory without spacetime might be `empirically incoherent'. That is, any theory which entails that the observations apparently supporting it are impossible, cannot receive empirical support \citep{bar96}---it undermines the very grounds for believing it. Since all observations ultimately involve events localized in spacetime, it might seem that theories without spacetime in their basic formulation are threatened with empirical incoherence; the confirmation of such a theory might be ruled out a priori. However, it is clear that a conflation is involved. (More) fundamental theories of QG are non-spatiotemporal in the sense that spatiotemporal structure is missing in their more furniture; but it is perfectly consistent to think that spacetime is present as an effective object, arising from the more fundamental ones. (And that observation events can be identified within effective spacetime.) That is, QG will not be empirically incoherent if the appearance of spacetime can be adequately explained, and of course that is exactly what our case studies aim to do.

Second, while this book concerns the idea of `emergent' spacetime as it arises in QG, one of our key concerns has already arisen in discussions of a different kind of spacetime emergence. The quantum mechanical wavefunction of $N$ particles is not a function in ordinary space, but of the positions of all the particles: $\Psi(x_1,y_1,z_1; x_2,y_2,z_2; \dots; x_N,y_N,z_N)$. Thus $\Psi$ lives in `configuration' space, in which there are three dimensions for each particle. \cite{Alb:96} has argued that we should take the wavefunction `seriously' as the ontology of the theory, and conclude that configuration space is more fundamental than regular space---that the three dimensions of experience are mere appearances of the $3N$ dimensions of reality. Whatever the merits of that view, the general idea has been attacked by Tim \citet{mau07}. In particular he argues as follows: one might

\begin{quote}
derive a physical structure with the form of local beables from a basic ontology that does not postulate them. This would allow the theory to make contact with evidence still at the level of local beables, but would also insist that, at a fundamental level, the local structure is not itself primitive. ... This approach turns critically on what such a derivation of something isomorphic to local structure would look like, \emph{where the derived structure deserves to be regarded as physically salient} (rather than merely mathematically definable). Until we know how to identify physically serious derivative structure, it is not clear how to implement this strategy. (3161)
\end{quote}

We have italicized the key phrase here. Suppose that one managed to show formally that certain derivative quantities in a non-spatiotemporal theory took on values corresponding to the values of classical spatiotemporal quantities; one would then be in a position to make predictions about derived space. However, according to the passage quoted, such a derivation (even if the predictions were correct) would not show that spacetime had been explained. In addition, we have to be assured that the formally derived structure is `\emph{physically salient}'. We agree with Maudlin that physical salience is required of proper---one can say `explanatory'---derivations: otherwise one simply has a formal, instrumental book-keeping of the phenomena. Indeed, we agree with him that the issue is particularly pressing in theories of emergent spacetime. But we think that it can be addressed in QG: one of the goals of this book is to investigate the (novel) principles of physical salience for theories of QG, the principles whose satisfaction makes the derivations of spacetime physically salient. In the following chapters we will look in detail at the derivations, to make clear the assumptions and forms of reasoning that lie behind them. In the concluding chapter 10 we will analyze what have learned, to start to explicate what makes a derivation of spacetime in QG physically salient.

But to explain that project---and its relevance to philosophy---we need to unpack the very notion of physical salience, as we understand it.\footnote{We are grateful to Maudlin for conversations on this topic. We believe that we capture the essence of his idea, even if we might differ in details; and especially regarding the depth of the problem in the case of spacetime emergence. We do, however, want to point out an important difference between the cases of emergence from QG and from configuration space: in the latter, but not the former, there is a way to formulate the theory in 3-space (as single particle wavefunctions with a tensor product). Thus in QM (but not QG) Maudlin can argue that the derivation isn't physically salient, because the formulation from which it is derived is unnecessary in the first place. That the crucial difference between QG on the one hand and QM (and GR) on the other lies in there being no alternatives translates into a different status for spacetime functionalism in the two cases has been argued by \citet{lamwut20}.}

\section{Physical salience}
\label{sec:introPS}

There is a subtlety about the way that Maudlin makes the point, however (which we did not clearly address in \cite{HugWut:13}). For the target derived structure in itself is prima facie physically salient: it is the physical datum to which the more fundamental theory is answerable. (Perhaps a more fundamental theory will show that some less fundamental theory is profoundly confused; but more generally one expects that existing, well-confirmed theories have latched onto some genuine physical structures, and that new, better theories will simply explain how, by subsuming the old in some broad sense.) So in that sense there is really no question of the physical salience of the `derived structure'---in the sense of the structure \emph{to be} derived. 

Rather, Maudlin is talking about a formal derivation within a proposed new theory, and the question of whether what is at present simply a mathematical structure, in numerical agreement with the target structure, in fact explains it, and isn't merely an instrument for generating predictions. We would break this question down into two interconnected parts (which will also help illuminate what is involved in explanation here). First, the question of whether and how the basal objects or structures of the more fundamental theory accurately represent physically salient objects and structures. As we shall see shortly, that question becomes far more pressing when none of the putative objects or structures are supposed to be in spacetime. Second, does the formal derivation of the phenomenal from the more fundamental make physical sense? That the derivation exists shows that it makes sense at the level of the formalism, and especially that the derivation is compatible with the mathematical laws. But, as Maudlin suggests, there is more to the question of physical salience than that. And the question is especially pointed when one wonders how the spatiotemporal could ever be `made' of the non-spatiotemporal. We will illustrate these ideas with a homely (and idealized in many ways) example.\footnote{The following has also been discussed in \cite{Hug:18}.}

The ideal gas law tells that for a gas (in a box of fixed volume) pressure $\propto$ temperature. Ideal gas theory says nothing about the microscopic composition of gases, so these are (among) the primitive quantities of the theory, operationalized via pressure gauges (relying on forces measured via Hooke's law for springs), and thermometers (so relying on the linear expansion with temperature of some substance). This is the phenomenon to be explained by the more fundamental theory, the kinetic gas model, according to which the gas is composed of atoms with mass $m$, whose degrees of freedom are their positions and velocities. The latter can be expressed by a vector $\vec V$, with 3$n$ components: for each of the $n$ atoms that make up the gas, three components, to describe the speed with respect to each of the three dimensions of space. Each atom has a kinetic energy associated with its velocity ($1/2m\vec v^2$); the average kinetic energy is simply their sum, divided by $n$: denote this quantity

\begin{equation}
T(\vec V)\equiv\overline{\frac{1}{2}m\vec V^2}.
\end{equation}

Now one computes the atoms' momentum change (per second per unit area) resulting from their collisions with the sides of the box: assuming that the collisions are elastic, and that the atoms are distributed evenly throughout the box and with respect to their velocities, one formally derives that 

\begin{equation}
P(\vec V) \equiv \frac{2n}{3v}\overline{\frac{1}{2}m\vec V^2}.
\end{equation}
 
Clearly the two quantities are proportional:

\begin{equation}
P(\vec V) \propto T(\vec V),
\end{equation}
which has the form of the ideal gas law. However (and despite the suggestive names, $P$ and $T$) we have so far said nothing to justify identifying the quantities with the pressure and temperature of the ideal gas law; we have only noted a formal proportionality. 

From this example we can abstract the following schema:

\begin{quote}
If \textbf{(a)} fundamental quantities $X$ can be `aggregated' into $\alpha(X)$ and $\beta(X)$, such that \textbf{(b)}  $f(\alpha(X))=g(\beta(X))$ follows from fundamental laws, then the law $f(A)=g(B)$ relating less fundamental quantities $A$ and $B$ is \emph{formally derived}.
\end{quote}
The term `aggregated' is supposed to be vague, in order to accommodate the many ways a derivation might proceed. But the underlying idea is that the more fundamental theory has (many) more degrees of freedom than the less fundamental, and somehow the more fundamental must be `summarized' by the less, for example by averaging, or by coarse-graining.

Maudlin's claim is that formal derivability does not suffice to properly derive phenomena: in particular, 3-dimensional space can be formally derived from the full $3N$-dimensional configuration space, but for Maudlin that does not make it a plausible, more fundamental alternative to ordinary space. And more generally, one should worry that a merely formal condition does not distinguish  instrumental calculi from serious physical accounts. And indeed, further analysis of the derivation of the ideal gas law shows that considerations of physical salience are at play.

In particular, $P(\vec V)$ is derived by assuming that the atoms are striking the sides of the box, and exerting a force there: so acting exactly at the place and in the way that would produce a reading on a pressure gauge. And $T(\vec V)$ is (according to the randomness assumption) the amount of energy in any macroscopic region of the box, say the location of the bulb of a thermometer: and collisions with the bulb will transfer kinetic energy to the molecules of the thermometer, causing thermal expansion. Imagine if instead that $P(\vec V)$ only referred to the center of the box; or if $T(\vec V)$ referred to a single atom in the box. Then the formal derivation would not be convincing. Or suppose that instead of the atomic gas model we imagined that a gas was a continuous object, whose degrees of freedom were somehow described by $\vec V$, but not as the velocities of anything (certainly not atoms). Then the formal consequences of kinetic gas theory could still be taken to hold, but they would no longer have the interpretation that they do in the kinetic gas model; the whole derivation would go through, but its physical meaning would be obscure. In short, the reason, in addition to their proportionality, that we find $P(\vec V)$ and $T(\vec V)$ convincing as pressure and temperature, and not just quantities following a similar law, is that they are spatiotemporally coincident with those quantities, and involve processes capable of producing the phenomena associated with those quantities. The derivation is not merely formal, but also physically salient.

Continuing our schema:

\begin{quote}
\noindent A (non-instrumental) derivation of phenomena requires, in addition to a formal derivation, that \textbf{(c)} the derivation have \emph{physical salience}.
\end{quote}

Of course, this schema does not tell us what it is to be physically salient, but the example of the ideal gas above illustrates two very important aspects, spatiotemporal coincidence, and the action of a physically accepted mechanism. And this observation immediately reveals the problem for the emergence of spacetime, because such criteria simply cannot be satisfied by derivations from non-spatiotemporal theories, because they are explicitly spatiotemporal criteria. For instance, it makes no fundamental sense in such a theory to even ask \textit{where} a structure is. So if such criteria are a priori constraints on science, then the QG program, to the extent that it involves non-spatiotemporal theories, is in some serious trouble. However, a second example indicates the contextuality of physical salience, and thereby the way in which QG can hope to achieve physical salience in its derivations.

\begin{figure}[htbp]
\begin{center}
\includegraphics[width=5in]{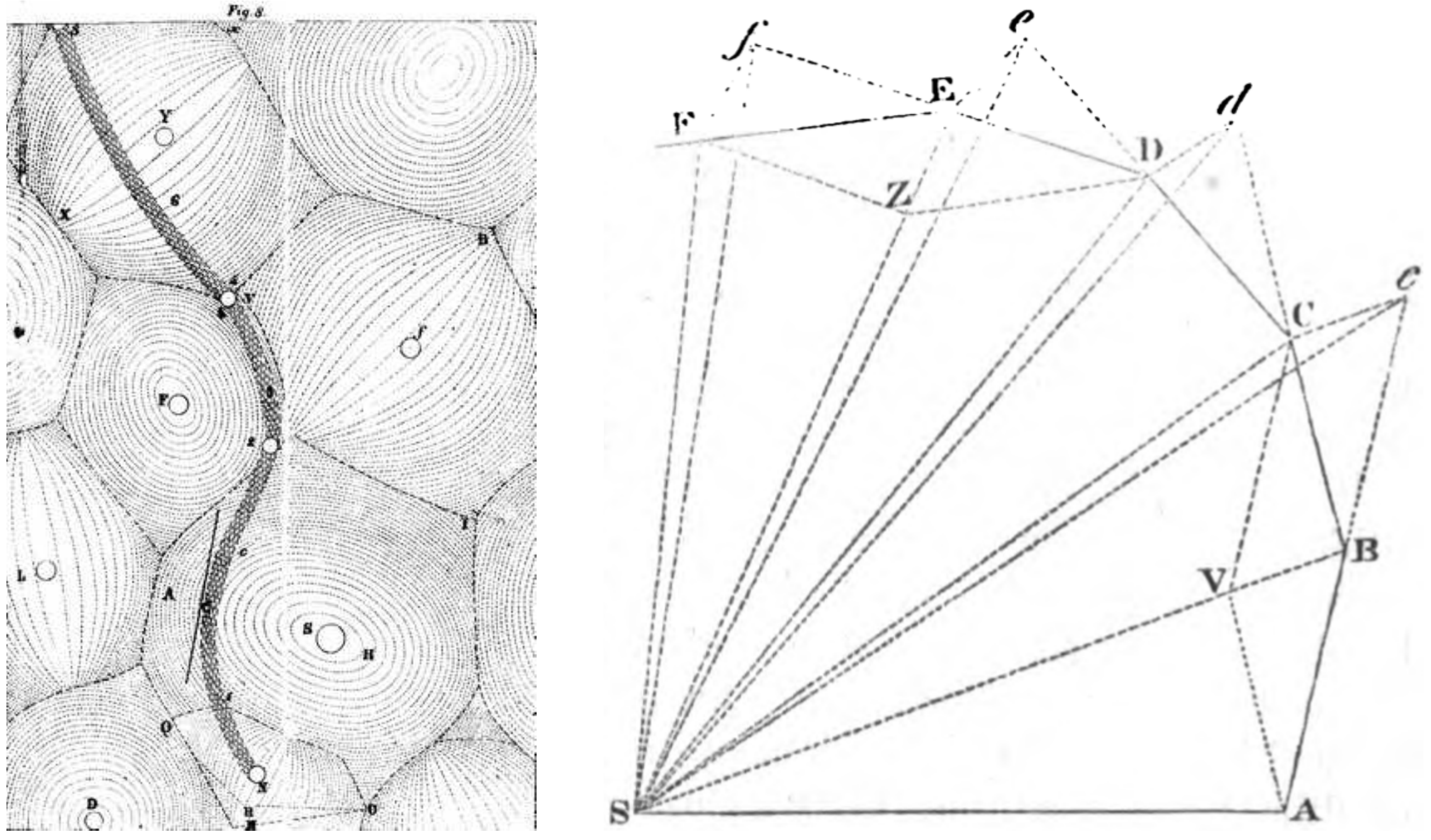}
\caption{Descartes' and Newton's competing images of gravity. On the left is pictured Descartes' vortex model: each cell represents a ball of rotating matter, with lines to indicate the direction of rotation (e.g., those surrounding $f,L,Y$ rotate about axes in the plane shown, while those surrounding $D,F,S$ rotate about axes perpendicular to the plane). The bodies at the center of a cell represent suns: $S$ is ours. On the right is the diagram from Newton's Proposition I.1 proof of Kepler's equal areas law for a central force (essentially, conservation of angular momentum). All that matters is the direction of the force (towards the point $S$), not any `hypothesis' about its nature. 
Ultimately Newton will apply the proposition to the case in which $S$ is our Sun. (Public domain, via Wikimedia Commons and Google Books.)}
\label{fig:D+N}
\end{center}
\end{figure}

Consider the competing Cartesian and Newtonian accounts of gravity, exemplified by illustrations from the \emph{Principles of Philosophy} \citep{Des:44} and the \emph{Mathematical Principles of Natural Philosophy} \citep{New:26}, respectively: see figure \ref{fig:D+N}. On the one hand we have the vortices of Descartes, which aimed to provide a mechanical account of gravity, in terms of the motions and collisions of particles. On the other there is Newtonian action at a distance, which allowed him (as in Proposition I.1) to formulate and use his mathematical principles. We pass over Newton's own ambiguous attitude towards the causes of gravitation (his refusal to `feign hypotheses' on the one hand, but his speculations in the \emph{Optiks} \citep{New:30} on the other). The point to which we draw attention is the controversy between the Newtonians and Cartesians regarding the need for mechanical explanation.\footnote{Note especially that we strictly distort the logic of Newton's \emph{Principia} here: as far as Proposition I.1 is concerned, the forces could be impulses directed towards the point $S$. However, though Newton's reader may not at that stage know the nature of the force, for Newton the figure represents the action of universal gravitation.} For the latter, Newton might have captured the effects of gravity in a formally accurate way, but offered no scientific explanation for the phenomena. For example, consider Leibniz's clear statement to Clarke:

\begin{quote}
If God would cause a body to move [round a] fixed centre, without any [created thing] acting upon it \dots it cannot be explained by the nature of bodies. For, a free body does naturally recede from a curve in the tangent. And therefore \dots the attraction of bodies \dots is a miraculous thing, since it cannot be explained by the nature of bodies. (\emph{Leibniz-Clarke Correspondence} in \citet{LeiClaLei:56})
\end{quote}

\noindent We take Leibniz's complaint to be exactly that Newton's derivation of the phenomena lacks physical salience, because only mechanical causes are physically salient explanations of unnatural motions. 

Of course, the Newtonians were ultimately victorious, and this Cartesian condition of physical salience was replaced by one that allows action at a distance, because of the success of universal gravity, and the failure of mechanical alternatives, such as Leibniz's. But that was not the end of the story: through the development and empirical success of electromagnetic theory, culminating in the development of special relativity, action at a distance was again rejected, with contact action replaced by the demand for local field interactions---and hence the replacement of Newtonian gravity with general relativity. Again, we understand this demand as a criterion of physical salience, required for more than merely formal accounts. But even that is not the end of the story, for quantum mechanics experimentally conflicts with that concept of locality, and so quantum non-locality must be accommodated in some way. (\citealt{Hes:61} is a classic telling of this tale.) 

By now, three points are indicated by this story: first, questions of physical salience, here in the form of the principles of locality, are genuine, controversial components of scientific enquiry. Second, such principles are historically contingent, changing in step with major advances in physics. Third, such changes are ultimately settled by, and epistemically justified by, empirical success: one of the things that we learn in a scientific revolution is a set of criteria of physical salience for explanation appropriate to the new domain of enquiry. Put this way, we see principles of physical salience as part of what Kuhn called the `disciplinary matrix' in the \emph{Postscript} to the second edition of (\citeyear{Kuh:62}), or what \cite{Fri:01} refers to as the `relative, constitutive a priori'. Though changes in the principles change wholesale what theories are even candidate explanations, we don't infer any catastrophic incommensurability here: as we said, innovations in physical salience are grounded in empirical success, like all other scientific knowledge. 

So we have a general answer to the problem raised earlier. \emph{How can a derivation of spacetime from a non-spatiotemporal theory ever be physically salient?} Well, it cannot satisfy the standards of physical salience that apply to theories with classical spacetime, but we should expect a non-spatiotemporal theory to require new standards. And so the real question is what are those new principles? Like Friedman, we see that question, and the development of such new principles as a foundational, interpretational, conceptual---hence \emph{philosophical}---endeavor. We shall elaborate on how such a project is to be conducted in \S\ref{sec:rolepp}. We will see throughout the book how this endeavor is ineliminably philosophical in the different approaches to QG. For now we want to illustrate the problem with an example.

\section{Non-commutative geometry}
\label{sec:introNCG}

Of necessity, this section is somewhat more technical than the others, and could be skipped by those not requiring a concrete illustration of how interpretational considerations come into play in elevating a formal derivation into one (potentially) having physical salience. It elaborates an example of a non-spatiotemporal theory already given, to show how one might come to view it as a theory from which spacetime emerges.

We start with familiar, commutative geometry, for which $xy=yx$, in a smooth manifold of points; let it be 2-dimensional for simplicity.\footnote{This section is based on \cite{HugLizMen:20}. See also \cite{Liz:09} for a more mathematical survey.} Consider polynomials $\mathcal{P}(x,y)$ of $x$ and $y$. These are `fields', meaning that they return a numerical value at each point $(x,y)$. They form an algebra with respect to multiplication: this just means that when you multiply two polynomials together, the result is another polynomial. Moreover, because $xy=yx$ we have that $\mathcal{P}(x,y)\mathcal{Q}(x,y)=\mathcal{Q}(x,y)\mathcal{P}(x,y)$, so that the algebra is commutative. (Check with $\mathcal{P}(x,y)=xy$ and $\mathcal{Q}(x,y)=x^2 +y^2$ if you like.) 

It may seem like a rather uninteresting structure, but in fact such algebraic relations alone contain geometric information about the space: in this case, that it is smooth, that it is 2-dimensional, and whether it is open or closed. This fact is shown by the important Gelfand-Naimark theorem (\citeyear{GelNai:43}), which is the foundation of `algebraic geometry'. Indeed, the whole structure of differential geometry can be recast in algebraic terms. (An interesting application is Geroch's (\citeyear{ger:72a}) formulation of general relativity as an `Einstein algebra'; discussed by \citet[\S9.9]{ear:89b} as a possible response to the hole argument.)

For a mathematician, the question of what happens when the algebra is `deformed' so that it is no longer commutative is irresistible: so one sets $xy-yx=i\theta$ and sees what happens. (And similarly in spaces of any dimensions.) Surprisingly, one finds that the structure necessary to cast geometry in algebraic terms remains (at bottom, one can still define a derivative on the algebra, in terms of which the other structure is defined). Moreover, the Euler-Langrange equation and Noether's theorem do not require commutativity, and so the structure of modern physics is preserved, even in such a `non-commutative spacetime'---in a purely algebraic formulation. 

But suppose such a physics were correct: how could it explain spacetime as it appears to us? Specifically, how are we to understand events localized in space in terms of an abstract algebra? When the algebra is commutative, the Gelfand-Naimark theorem lets us interpret the elements as fields, $\mathcal{P}(x,y)$ related to regions of space; but what about the non-commutative case? The question is just that which has concerned us in this chapter (and indeed the whole book): how can we derive the appearance of classical spacetime from a non-spatiotemporal theory, in a physically salient way?

The obvious thing to try is to (a) interpret $x$ and $y$ not as elements of an abstract algebra, but as fields in an ordinary plane: taking the value of the $x$ and $y$ coordinates at any point $(x,y)$. Then (b) define a new binary operation, $\star$, such that $x\star y-y\star x=i\theta$. Then (c) construct the algebra of polynomial fields, but with $\star$-multiplication instead of regular (point-wise) multiplication. Indeed, this is exactly how one typically proceeds in non-commutative geometry: in one formulation, the operation is `Moyal-$\star$' multiplication\footnote{$\phi\star \psi \equiv \phi\cdot\psi\ + \sum^\infty_{n=1}(\frac{i}{2})^n\frac{1}{n!}\theta^{i_1j_1}\dots\theta^{i_nj_n}\partial_{i_1}\dots\partial_{i_n}\phi\cdot\partial_{j_1}\dots\partial_{j_n}\psi$.}, and the fields form the `Weyl representation' of the algebra. 

The algebra of the fields with respect to $\star$ will be that of the abstract non-commutative algebra, and now we have referred that algebra to objects in an ordinary manifold. In particular, one could talk about the local region in which such-and-such a field has values less than 1, say. Indeed, one might now wonder whether we should throw away the abstract algebra, and just treat physics in `non-commutative geometry' as really physics in commutative geometry, but with an unfamiliar multiplication operation. In other words, wonder whether classical spacetime needs to be recovered at all? 

Huggett, Lizzi, and Menon (\citeyear{HugLizMen:20}) argue that indeed it must be, for the Weyl representation has formal representational structure that exceeds its meaningful, physical content. In particular, the concept of a region with an area smaller than $\theta$---a forteriori that of a point---is undefinable in the theory. This can be seen in a couple of ways, but for instance the attempt to measure positions more accurately leads to unphysical results. The conclusion is that, although the Weyl representation contains points and arbitrarily small regions, they are purely formal, and do not represent anything real: non-commutative geometry---even the Weyl representation---is physically `pointless'.

As a result, we cannot understand a point value of the Weyl fields as having any physical meaning. Rather we need to understand the fields as complete configurations: the unit of physical meaning for a field in non-commutative space is the \emph{function} from each point to a value, $\mathcal{P}:(x,y)\to\mathbb{R}$; not its \emph{value}, $\mathcal{P}(\mathrm{x,y})$ at any particular point $(\mathrm{x,y})$. But the full configuration is equivalent to the place of the field in the abstract algebra, and so we are back to the question of deriving locality.

Here is one way to proceed, using an ansatz proposed by Chaichian, Demichev, and Presnajder \citeyear{ChaDemPre:00} (discussed further in \citealt{HugLizMen:20}). They propose that an ordinary, commuting field---the kind observed in classical spacetime---be related to a Weyl field $W(x,y)$ by an operation of `smearing'. One multiplies $W(x,y)$ by a $\theta$-sized `bell function' about $(X,Y)$, and integrates over the Weyl space coordinates $x$ and $y$.\footnote{$\Omega(X,Y) \propto \int \big(e^{-((X-x)^2+(Y-y)^2)/\theta}\cdot W(x,y)\big)\ \mathrm{d}x\mathrm{d}y$.} The result is a new field $\Omega(X,Y)$. Extrapolating from this `CDP ansatz', the result of smearing is to introduce classical space into the theory. $W$ lives in Weyl space, whose status, we argue, is only that of a formal representation of the fundamental algebra, while $\Omega$ should be interpreted as living in the physical space that we observe. We thus interpret smearing as relating a \emph{function on one space} to a \emph{value on another space}: it relates the non-commuting field $W$, represented as a function over Weyl space points $(x,y)$, to the \emph{value} of an observed, commuting field $\Omega$ at physical space point $(X,Y)$. That it takes a function to a value is just mathematics; that it relates Weyl and physical spaces is a substantive physical postulate.

However, it still makes no sense to consider $\Omega$ in regions smaller than $\theta$: we have in fact erased the unphysical information at such scales by smearing $W$. So strictly a single coordinate pair $(X,Y)$ does not label a physical point. Rather, the proposal is that these smeared fields are approximated by observable fields over regions greater than $\theta$; thereby formally deriving the latter, spatially localized objects from the former, purely algebraic objects. (In this case, we have smearing as `aggregating', in a very loose sense.) Of course, in our existing theories, fields live in a full commuting spacetime, but that is an extrapolation from our actual observations of fields, which are always over finite regions, to date larger than $\theta$. On the proposed interpretation, then, any information contained in $\Omega$ about regions less than $\theta$ is not only unobserved, but unphysical, surplus representational `fluff'. 

Now, nothing in the theory forces this picture as a physical story---it is merely an interpretational postulate. (Though we claim that it is conceptually coherent.) However, it has empirical consequences: the dynamics magnifies the $\theta$-scale non-commutativity to observable scales (e.g., \citealt{carhar+:a}). If those predictions are successful, then we have evidence that the underlying non-commutative field theory \emph{and} the interpretational postulate are correct. Imagining that situation then, we claim that the situation is exactly analogous to that of the Newtonians regarding action at a distance. That is, we would be justified in accepting the CBP ansatz and our interpretational postulate as novel principles of physical salience: they regulate what constitutes a physically salient derivation in the theory. In both cases, the final ground is the empirical success of the theory.

So it should be clear how the example illustrates our points about physical salience and our scheme proposed above. In the first place we have argued that non-commutative geometry is non-spatial, in the sense that it is `pointless', and so must be understood as a purely algebraic theory. Then we have explicated a possible formal derivation of localizable fields from this more fundamental theory. And finally, we have sketched a scenario in which such a derivation leads to successful predictions, and hence to the conclusion that the formal derivation is physically salient, in fact \emph{explaining} the appearance of localized fields, and ultimately classical spacetime. Of course, we highlight that this discovery constitutes a \emph{change} in what derivations `deserve to be regarded as physically salient (rather than merely mathematically definable)', to paraphrase Maudlin. 

This is the pattern that we will see in more detail in the examples of the following chapters, and to which we will return in the conclusion. In the following two sections we will investigate further what is achieved by such a derivation (\S\ref{sec:funcES}), and use our account of physical salience to indicate what philosophy should aim to do when engaging emerging physics such as QG (\S\ref{sec:rolepp}).

\section{Spacetime functionalism}
\label{sec:funcES}

The schema that we presented for a physically salient derivation relates closely (but not exactly) to Lewis' (\citeyear{Lew:72}) account of functional identification. Since `spacetime functionalism', of various varieties, has been recently discussed it is worth drawing the comparison, to better understand the nature of the proposed emergence. Suppose, in idealization, that a theory $T$ is formulated as a postulate `$T[t]$'. $t$ represents what was traditionally called the `theoretical' terms, though we prefer `troublesome' \cite[\S3.1]{WalBut:18}: the idea is that these are the new terms introduced by the theory, and the `trouble' is the question of how they garner meaning. `$T[\cdot]$' also involves (traditionally) `observational', or (according to Lewis) `old', or (with Walsh and Button) `okay' terms, and Lewis proposes that the $t$ are defined in terms of them by 

\begin{equation}
\label{eq:fdef}
t=\rotatebox[origin=c]{180}{$\iota$}x\ T[x].
\end{equation}
That is, `the $t$ are (if anything) the extant, unique things that satisfy the theory postulate'. The $t$ are thus defined in terms of their nomic relations to one another and to the okay terms---i.e., in terms of their `functional' relations---and so (\ref{eq:fdef}) is a functional definition.\footnote{Lewis proposes in passing that the actual definition be modified to allow for \emph{approximate} satisfaction of $T[\cdot]$. In our opinion that is always going to be the case in actual theories, so this modification is not optional but required, and his discussion is a significant idealization. The harder question of how exactly the modification is to be implemented is not carefully addressed by Lewis.} As a result, the terms $t$ are rendered semantically okay (though they may remain metaphysically problematic).

Suppose that we also hold a postulate $R[r]$, where $r$ and $t$ do not overlap, so that our acceptance $R$ does not depend on our views on the troublesome terms of $T$.\footnote{Acceptance of a theory requires that it be meaningful, hence acceptance of $R[r]$ requires that the $r$ be referential, and thus that any troublesome $r$ can be functionally defined in terms of the okay $r$ as in (\ref{eq:fdef}), mutatis mutandis. We accept this assumption for Lewis' cases, but we will see that things are more complex in the case of QG.} Further suppose that we come to believe $T[r]$. Now,

\begin{equation}
T[r],\ t=\rotatebox[origin=c]{180}{$\iota$}x\ T[x]\ \vDash\ r=t,
\end{equation}
so, by definition of the $t$, $T[r]$ deductively entails the identity of the objects of $R$ and $T$. Lewis' point is the epistemic one that such functional identifications are thus not inductive: given a functional definition, once one accepts that the objects of $R$ play the same roles as those of $T$, \emph{logic and meaning alone commit one to accepting their identity}.

Lewis offered electromagnetic waves and light as an example of the scheme, but of course his point was that `when' neuroscience showed that neural states played the functional roles of mental states, then they would---as a matter of logic and definition---be identified. The subject of this book also broadly fits Lewis' scheme:  theory $T$ is our spacetime theory assumed by QFT and relativity theory, while $R$ is a theory of QG. Denote them $ST$ and $QG$, respectively, and use $ST$ to functionally define any troublesome spacetime terms. Then, according to the schema of \S\ref{sec:introPS}, a physically salient derivation of $ST$ from $QG$ shows that `aggregates' of $QG$, described using its terms $q$, satisfy $ST[\cdot]$---that they indeed play the functional roles of the objects of $ST$. So the identity follows. However, there are differences to Lewis' functionalism, which we will explain presently.

Now, how to turn Lewis' scheme into a concrete plan for the functional reduction of spacetime? The spacetime functionalism recently introduced by \citet{lamwut18,lamwut20} is based on the general scheme in the spirit of \citet[101f]{kim05} according to which a functional reduction of higher-level properties or entities to lower-level properties or entities consists in two necessary and jointly sufficient steps:\footnote{Kim's model involves three steps, where the second is to identify the entities in the reduction base that perform the role at stake, and the third is to construct a theory explaining how these fundamental entities perform that role. We subsume these two steps in our second stage.} 

\begin{enumerate}
\item[(FR1)] The higher-level entities/properties/states to be reduced are `functionalized'; i.e., one specifies the causal roles that identify them, effectively making (\ref{eq:fdef}) explicit.
\item[(FR2)] An explanation is given of how the lower-level entities/properties/states fill this functional role, so that we come to accept $T[r]$.
\end{enumerate}
If these two steps are fulfilled, then it follows that the higher-level entities/properties/states are realized by the lower level ones.

Applying the template of functional reduction to the case of the emergence of spacetime in QG, the two steps above become: 
\begin{enumerate}
\item[(SF1)] Spacetime entities/properties/states, $s$, are functionalized by specifying their identifying roles, such as spacetime localization, dimensionality, interval, etc. Effectively, one makes explicit $s=\rotatebox[origin=c]{180}{$\iota$}x\ ST[x]$.
\item[(SF2)] An explanation is given of how the fundamental entities/properties/states, $q$, postulated by the theory of quantum gravity fill these roles, so that we come to accept $ST[q]$.
\end{enumerate}
Again, if these two steps are fulfilled, it follows that the (perhaps aggregated) QG entities/proper\-ties/states \emph{are} the spacetime entities/properties/states. In the following chapters, after explaining each theory of QG and its conceptual foundations, we will follow this scheme in our discussions: 
on the one hand describing the functional roles of spacetime entities/properties/states, and on the other showing how the theory of QG proposes that those roles are played. Of course, given the evidential state of QG, we do not claim that that these proposals are correct: we only describe how the theories \emph{may} functionally reduce spacetime, not---as far as we currently know--how they \emph{do}.

Several remarks are in order regarding the functionalist approach to spacetime. First, it should be made clear that we take emergence and reduction to be compatible with one another, and hence functional \emph{reduction}  
may serve as a template to explain the \emph{emergence} of a higher-level feature, i.e., the fact that higher-level entities exhibit novel and robust behaviour not encountered or anticipated at the more fundamental level.\footnote{As restated many times in our earlier publications, and in agreement with what we take to be the consensus in philosophy of physics as stated, e.g., in \citet{but11a,but11b} and \citet[\S2]{cro16}.}

Second, there is a sense in which a functionalism about spacetime must start from a broader conception of functional reduction than is usual in the familiar functionalisms in the philosophy of mind or the philosophy of the special sciences. There, a mental or biological or other higher-level property is understood to be determined by---indeed, usually \emph{identified with}---its \emph{causal role} within the relevant network such as the network of mental or biological activities. If in spacetime functionalism the roles are still supposed to be \emph{causal}, then a much broader notion of `causal' must be at work, one that does not in any way depend on the prior existence of spacetime. As it is not clear what that would be, it is preferable to formulate a notion of functionalism devoid of any insistence that the functional roles be causal. 

Third, the central claim of spacetime functionalism is that it is sufficient to establish only the functionally relevant aspects of spacetime. In particular, it is therefore not necessary to somehow derive relativistic spacetime in its full glory and in its every aspect in order to discharge the task. Naturally, this raises the question of what these functionally relevant aspects of spacetime are---the task of (SF1). As we will see in the following chapters, different approaches to QG take different stances on what functions are to be recovered, though broadly speaking, all aim to recover functions sufficient for the empirical significance of basic metrical and topological properties. Our stance will be that the list of functions cannot be determined \emph{a priori} from conceptual analysis of classical spacetime theories, but by the twin demands of the empirical, and of the resources of the proposed reducing theory. In short, part of the work of each chapter will be to identify the spacetime functions recovered in the different approaches, and indicate how they relate to observation.

Fourth, the scheme permits a form of `multiple realizability', as is typical of functionalism also in the philosophy of mind or the philosophy of the special sciences: (SF2) allows that different (kinds of) fundamental entities might play one and the same functional role, i.e., that the `realizer' of spacetime might have been by something other than what it in fact is. This liberal stance spurs a concern that functionalism is too weak a condition to secure the emergence of spacetime, that the true nature of spacetime is not exhausted by its functional roles, so that none of the mere functional realizers could ever truly be spacetime. In particular, the worry continues, a rash reliance on functionalism misses the qualitative nature of spacetime---some kind of spacetime `qualia', as it were---and it is precisely such qualitative features that make spacetime what it is, and which cannot be recovered by mere functional realization. However, the case of spacetime is disanalogous to that of mind: we agree with \citet{knox14} who states that where ``the fan of qualia [in the philosophy of mind] has introspection, the fan of the [spacetime] container has only metaphor'' (16), and with \citet{lamwut18} who agree that the ``nature and status of the evidence in favour of [mental] qualia may be equivocal, but the alleged ineliminable intrinsically spatiotemporal but ineffable quality of spacetime substance remains positively elusive'' (43f). We conclude with them that the qualia worry in this form gets little if any traction in the spacetime case.\footnote{We also concur with \citet{lamwut18} in their rejection of the version of this concern articulated in \citet{ney15}, who worries that if the fundamental entities are not already appropriately (spatio)temporal in their nature, they cannot `build up' or constitute spacetime as they are not the right kind of stuff (see also \citealt{hagar2013primacy}). As diagnosed by Lam and W\"uthrich, advocates of this worry seem to rely on an unreasonably narrow concept of constitution. We might also object that if we surrendered to this worry, there would be no principled reason to think that it would not also annihilate all other cases of presumed emergence and amount to an unyielding dualism.} 

Borrowing a distinction from \citet{blb18} between a ``hard'' and an ``easy problem'' of spacetime emergence, spacetime functionalism amounts to the denial that there is a hard problem of an unbridgeable explanatory gap between the fundamental, non-spatiotemporal and the emergent, spatiotemporal realm. For the functionalist, what is to be shown---by a physically salient derivation---is how the fundamental degrees of freedom can collectively behave in ways such that they play the required spacetime roles. And nothing more. No special character, or essence, or metaphysical nature need be accounted for. Functional identification requires no `luminosity' of light beyond the behavior of electromagnetic waves, or `consciousness' beyond the functioning of neurons. Or in our case, no special `spatiotemporality' that the non-spatiotemporal could never obtain. Once one has shown that the non-spatiotemporal plays the roles of the spatiotemporal---and so \emph{is} the spatiotemporal---no more need be said: one has a full scientific account of the emergence of spacetime, and no `explanatory gap' remains.

Fifth, functionalism shows how the goal of reduction can be the scientific explanation of the functional roles of higher level entities/structures/states by lower level entities/structures/states (and to nothing more). But, it is debatable to what exactly spacetime functionalism is ontologically committed: substances, relations, entities, structures, states, or something else. We will not further pursue this debate as we believe it to be orthogonal to the concerns of this book. Thus we hope that the reader will forgive our switching between speaking of the spatiotemporal as if it were an entity, or a structure, or a state, or something different yet again. We simply aim to avoid torturing English more than necessary, and no deep philosophical commitment should, for instance, be read into our using `spacetime' as a noun.

Sixth, we are far from the first to suggest functionally defining space or spacetime. \citet[chapter 2]{DiS:06} reads Newton's Scholium to the definition in much this way (though \citealt{Hug:12} disagrees). Functionalist strategies have also become very visible in the philosophy of non-relativistic quantum mechanics, where \citet{wal:12} deploys it in his defence of an Everettian interpretation and \citet[Ch.\ 6]{alb15} in support of wave function monism. Those latter applications differ from ours because they are concerned with recovering three-dimensional physical space. In contrast, spacetime functionalism in QG is commissioned with functionally recovering  4-dimensional spacetime, and so relates to work by \citet{knox13,knox14,knox19} in the context of classical spacetime physics. For her, something `plays spacetime's role' and thus \emph{is} spacetime ``just in case it describes the structure of inertial frames, and the coordinate systems associated with these'' (\citeyear[15]{knox14}). In GR, the metric field performs spacetime's role in this sense and thus is identified with spacetime by her. As the metric may itself not be fundamental but instead emerge from the collective behavior of more fundamental degrees of freedom, she explicitly leaves open the possibility that the realizers of spacetime's functions may themselves not be fundamental \citep[18]{knox13}. As the relationship between the fundamental degrees of freedom and the emergent spacetime realizer is left untouched by Knox's inertial frame functionalism, the latter does also not shed any light on it.\footnote{Cf.\ \citet[40]{lamwut18} and \citet[\S3]{lamwut20} for a more detailed discussion of inertial frame functionalism and how it relates to our project.} 

Seventh and finally, there is an important but subtle difference in the application of Lewis's scheme to QG from that in the cases he has in mind. Suppose we accept a spacetime theory $ST[s]$, where whatever it is that performs the spacetime functions is denoted by the troublesome terms, $s$. These, following Lewis, we take to be defined by

\begin{equation}
\label{eq:defST}
s=\rotatebox[origin=c]{180}{$\iota$}x\ ST[x].
\end{equation}
The okay terms appearing in $ST[\cdot]$ would refer to matter of various kinds, its relative motions and point-coincidences: so, for instance, the metric in GR might be defined locally in terms of its role in determining motions under gravity or scattering amplitudes. We think that this part of Lewis' picture---which corresponds to (SF1)---fits our cases well. But what about the second part of his scheme, involving $R[r]$? Although the result is still a functional identification, its significance has shifted somewhat, as we shall now explain.

Butterfield and Gomes (\citeyear{ButGom:20,GomBut:20}) analyze recent proposals for spacetime functionalism in explicitly Lewisian terms. They emphasize, as we have, that in Lewis' scheme theoretical identification follows by definition alone (once the $r$s are known to play the role of the $t$s), and that functional identification is a species of reduction. But they also show how various spacetime and temporal functionalisms follow the `Canberra plan', according to which the troublesome $t$s are not only defined by $T$, but are also `vindicated' by their functional identification as $r$s. For instance, as mental states, perhaps, turn out to be neural states so, in their examples, a temporal metric might be identified with purely spatial structure; then, if neural states or spatial structures are on a firm (or firmer) ontological footing than mental states or time, the identifications show that the latter are equally well grounded. They are, that is, vindicated against any metaphysical suspicions raised against them. That vindication is not by itself achieved by the functional definition (\ref{eq:fdef}) of the $t$s; that merely makes the terms referential, so that they can be meaningfully employed. Put another way, (FR1) alone does not vindicate the mental, for instance; (FR2) is also needed, to show how the mental is part of the physical.\footnote{Or put yet another way, the $t$ are often troublesome both semantically and ontologically: the functional definition takes care of the first problem, while the functional identification takes care of the second. When we use `troublesome' we always mean semantically.}  Regarding these cases, we are in agreement with Butterfield and Gomes emphasis of this important distinction, and its applicability to the cases that they discuss.

However, in our cases, for which $R$ is some $QG$, the second step, while still involving a functional identification, does not follow the Canberra plan, because the troublesome terms, $q$, of $QG$ are non-spatiotemporal, and so on a \emph{weaker}, not firmer, footing---the ontological and semantic correlate of empirical incoherence.\footnote{Butterfield and Gomes do not claim otherwise, and indeed acknowledge that QG will look different (\citeyear[3]{GomBut:20}).} Ontologically, as we have discussed, our physical and metaphysical categories assume spatiotemporality, and so the natures of the $q$ are mysterious. Semantically, we can expect an attempt to functionally define the $q$s as $q=\rotatebox[origin=c]{180}{$\iota$}x\ QG[x]$ to fail. Lewis' scheme for functional definition requires that a theory have sufficient okay terms to \emph{uniquely} define the troublesome ones: if many collections of terms satisfy the putative definition, then it fails to establish reference. But that is what one expects in a theory that breaks from established categories as radically as a non-spatiotemporal one; the terms that we take to be okay are systematically spatiotemporal in some way, and so are expected not to appear in $QG$. And indeed, we contend that the theoretical concepts of the theories we consider in this book cannot be defined without appeal to spatiotemporal concepts external to the basic formulation of the theory.

Given this situation, the significance of functional reduction is different from the way in which Lewis (and Kim) proposed. Rather than following the Canberra plan of vindicating spacetime objects by reduction, in our approach to QG things are \emph{reversed}: the non-spatiotemporal objects of $QG$ are vindicated via their identifications with spatiotemporal objects. Clearly this approach only works to the extent that the spatiotemporal is itself on a firm ontological footing, which of course is a topic of endless debate. To skirt such debates in this book we will remain as neutral as possible, and not take any stand on the metaphysical nature of spacetime features such as topology or metricity, so that our conclusions remain valid for anyone who accepts them under whatever interpretation.

Within Lewis' framework, the vindication of the $q$ works as follows. Suppose that non-spatiotemporal $QG[q]$ has been proposed. As explained, the $q$ are semantically troublesome and ontologically suspect. Moreover, until we accept that a derivation of (at least a fragment of) $ST$ is physically salient, we have no empirical grounds for accepting $QG$. Such a derivation will provide, not only grounds for $QG$, but also define and vindicate the $q$. Introducing the `aggregate operator' $\alpha(\cdot)$, according to our schema, when we have a physically salient derivation of spacetime properties, then we accept

\begin{equation}
\label{eq:fdefa}
\alpha(q)=\rotatebox[origin=c]{180}{$\iota$}x\ ST[x].
\end{equation}
In conjunction with $ST[s]$ this entails that 

\begin{equation}
\alpha(q)=s
\end{equation}
more-or-less as for Lewis. However, the reversal of the Canberra plan makes several things different. 

First, semantics. As noted, the $q$ were not antecedently defined, but now can be through their---or rather the $\alpha(q)$'s---role as spacetime entities/structures/states. In other words, (\ref{eq:fdefa}) is in part definitional of the $q$: the troublesome non-spatiotemporal terms of $QG$ can only be defined with reference to spatiotemporal terms not native in $QG$. Moreover, (\ref{eq:fdefa}) only succeeds in defining the $q$ if in physical fact they play the ascribed roles, and do merely mimic them formally; something that the physical salience of the derivation will secure.\footnote{(\ref{eq:fdefa}) is not purely definitional, since it also also involves an existential commitment that the $q$s exist. And it need not fully define $q$; we also still have that $q=\rotatebox[origin=c]{180}{$\iota$}x\ QG[q]$ by definition.} Second, ontology. The $q$s are placed on a firm ontological footing---are vindicated---when we accept that the $\alpha(q)$ are in physical fact those entities/structures/states that play the spacetime role.\footnote{In \citet[284]{HugWut:13}, we described this approach to vindication as physical salience flowing down to the $q$ `from above'.} Once again, acceptance of the physical salience of the derivation secures just that. 

Finally, epistemology. In Lewis' scheme, we have independently accepted theories of, say, neuronal and mental states, and \emph{later} discover that they play the same functional roles, entailing that they are identical. In our case, the acceptance that $QG$'s objects (or rather their aggregates) play the same roles as $ST$'s objects, and hence are identical with them, is \emph{simultaneous} with our acceptance of $QG$. In general terms, the evidence for $R[r]$ is no longer antecedent (or independent) of the evidence for $T[r]$, but rather the very same evidence. As such the epistemic calculus is different. In one case, observations of neuronal states can be made independently of mental states, and we only have to show that they perform the relevant functions: producing suitable behaviors, for instance. In the other, observations are not independent of spacetime states, and have to support  both the truth of a theory of QG, and that its objects perform the right functions. To give evidence, that is, that the formal derivation of those functions is indeed physically salient. As a result, although the deductive logic is the same, the empirical inference to the premises of the identification is different, and indeed weaker. However, as we say, it is of the normal empirical kind, and we fully expect it to be made for a successful theory of QG. There is no special ground for skepticism.\\

So much for the functionalism that lies behind the investigations of this book. But why is finding a functional reduction in any way a philosophical task, rather than one for physics?

\section{The role of philosophy in physics}
\label{sec:rolepp}

As we noted, the theories that we plan to investigate are all speculative at present, faced with considerable formal and empirical uncertainties. So what can we hope to learn from a philosophical enquiry into something that is at worst likely false, or at best a work in progress? We see the situation as characteristic of emerging fundamental physics (and perhaps other sciences). The process of discovery takes place along various fronts: obviously, new empirical work constrains theory and requires explanation; also obviously, new mathematical formalisms are tried out and explored; less obviously, but just as importantly, conceptual analysis of the emerging theory is undertaken. In particular, we want to stress that this last kind of work is carried out concurrently with the empirical and theoretical. One should not view interpretation as something that merely happens after an uninterpreted formal structure is presented, but as an inextricable aspect of the process of discovery. As such, it is something that has to be carried out on inchoate theories, in order to help their development into a finished product.

We claim that this view is supported by the historical record: we have in fact already seen this for Newtonian gravity. But one can equally well point to the absolute-relative debate in the development of the concept of motion, or 19th century efforts to come to grips with the physical significance of non-Euclidean geometry. These debates did not wait until after a theory was developed to clarify its concepts; rather they had to be carried out simultaneously, as an integral part of the development of the theory (see \citet{DiS:06}). Of course we are hardly the first to realize that such philosophical issues have to be addressed together with the empirical and theoretical ones. Many of Kuhn's (\citeyear{Kuh:62}) arguments illustrate this point, and more recently it is a major theme of \cite{Fri:01}. But while we agree with their focus on philosophical, conceptual analysis as an essential part of theory construction, we don't intend to get involved in issues involving the \textit{a priori} or incommensurability, instead we want to emphasize the practical role for analysis in the development of QG. 

In the search for a new fundamental theory, the goal is---as it was for Einstein and for Newton---a new formalism plus an interpretation that connects parts of the formalism to antecedently understood aspects of the physical world, especially to the empirical realm. That is, an interpretation of how the more fundamental plays the functional roles of the less fundamental. And of course that means undertaking the project that we have been talking about in this section, of deriving spatiotemporal predictions from theories of QG. But one never simply co-opts or invents formalism without some eye on the question of how it represents existing physics of interest; and as the formalism is developed it becomes possible to see more clearly how and what the new formalism represents. Addressing this question is of on-going importance for finding the right formalism for the area under study. Moreover, constructing such a formalism does not typically proceed in a monolithic fashion; instead different fragments of theory are proposed, investigated, developed or abandoned. For example, think of the development of the standard model of QFT from the early days of quantum mechanics. So the analysis of concepts of the new theory in terms of existing physics is often faced with a range of half-baked theories and models. All the same, lessons about how a more developed, less fragmented theory can be found depend on asking how the fragments represent known physics---the answers are potential clues to how the finished product could do so.

We believe that contemporary QG should be thought of in just this way---certainly the fragmentation is real! Our primary goal is to look at a range of the half-baked fragments and ask how they connect to spatiotemporal phenomena. Since they do not do so in a familiar way, in terms of a continuous manifold of points, the question becomes `how does spacetime emerge from the underlying physics?'. We hope, therefore, that by concentrating on the question of emergence, aside from all the other issues involved in the search for a theory of QG, we will be performing a service to physicists working in QG, by focussing their attention on what is already known---and reminding them that success depends on making it part of the search. Naturally, we do not expect to find solutions of the order of Newton or Einstein! Indeed, a lot of what we shall do is draw out answers already given by physicists; we believe that careful philosophical analysis of these answers can help clarify them to reveal strengths and weaknesses, and hence aid progress. (Moreover, because we are focussed on this quite narrow issue, we can survey a wider range of approaches than most physicists actively study, and so provide a helpful overview of the topic.) And hence we believe that in the examples we will consider there are important clues for the development of QG which philosophical analysis can reveal.

\section{The plan for the book}
\label{sec:finalintro}

Thus, our primary aim is to see how spacetime disappears and re-emerges in several approaches to QG, and to show how this is not just a technical issue for physicists to solve, but instead elicits numerous foundational and philosophical problems. As we work through three such approaches---causal set theory (CST), loop quantum gravity (LQG), and string theory, which were all briefly introduced in \S\ref{sec:WwoST}---, we bring these philosophical issues to the fore and will concentrate our discussion on them. 

It is common to divide approaches to QG into those which start out from GR and attempt to convert it into a quantum theory of gravity in different ways and into those departing from the standard model of particle physics and aim to add gravity to the other three forces of the standard model. In the former approaches such as CST and LQG, we would not expect the resulting theories to fold in the physics of the standard model, whereas the latter, such as string theory, will presumably deliver more encompassing, unifying theories. It is clear that either way, a theory of QG needs to address how the geometrical degrees of freedom of spacetime interact with the matter degrees of freedom present in the world. But it is also clear that both kinds of degrees of freedom may well look very differently from what we are used to from other theories.

The first two chapters after this one focus on CST. Chapter 2 introduced the basic kinematic axiom of the theory and shows how in it at least space disappears rather radically from the fundamental ontology, but also that temporal aspects do not all survive. This raises the immediate question of the relationship between the fundamental ontology of causal sets with that of relativistic spacetimes, a question we start to address in chapter 2. Although some functions of space can tentatively be recovered, what is needed is a more systematic understanding of how causal sets generically give rise to worlds which appear to be spatiotemporal in ways described, to good approximation, by GR. The way in which this `derivation' of spacetime is attempted in CST is sketched and discussed in chapter 3. In this chapter, we will discuss the role played by introducing a dynamics for the theory. We will argue that the emergence of spacetime in CST is closely tied to deeply philosophical questions regarding the metaphysics of space and time.


Chapter 4 and 5 turn to LQG, retracing the disappearance and emergence of spacetime in this approach. Just as CST, LQG builds a research program around what it takes to be GR's central lesson. In the case of LQG, this is the insight that GR postulates a truly dynamical spacetime, interacting with other fields. The demand is encoded in the theory's general covariance. LQG seeks to articulate a theory of QG by delicately applying known quantization procedures to a Hamiltonian formulation of GR. Chapter 4 chronicles and discusses whether and, if so, how this approach leads to the disappearance of spacetime. Unlike CST, it is time whose existence is much threatened in LQG than space. Chapter 5 seeks to understand how relativistic spacetime then emerges from the fundamental theory, finding, again, close ties to philosophical questions.

Other approaches apply the strategies of perturbative QFT---so successful in understanding the other forces---to quantize gravity. The technique calls for starting a system in which the fields do not interact to build up a space of states: a lowest, vacuum state, and states of discrete, particle-like `quanta'. Generally such a system is solved exactly, and the vacuum describes an obvious classical state. Then one introduces a small interaction, and uses approximation techniques to study the behavior of fields: especially the scattering of quanta. This approach was applied to gravity early on: Minkowski spacetime is a natural vacuum, and the gravitational field has quanta known as `gravitons', very analogous to photons, the quanta of the electromagnetic field. Indeed, quite a lot is known about the quantized gravitational field through such methods, and this knowledge is taken as a constraint on a successful theory of QG. However, divergences prevent the theory from being generally applied; moreover, these divergence cannot be adequately resolved by `renormalization' as they can for other QFTs.\footnote{See \citet[chapter 2]{Kie:12} for a very nice survey of QFT of the gravitational field.} 

String theory works within this approach, but with one important tweak: instead of quantized point like particles, it deals in quantized 1-dimensional, string-like objects. This, it appears, makes all the difference to the finiteness of the theory. Chapters 6-9, address the emergence of spacetime in string theory. Chapter 6 is a fairly technical introduction of the theory, aimed at philosophers of physics: it aims to be more intuitive, and more explicit about the conceptual and physical framework than physics textbooks usually are. For those who have some familiarity with classical and quantum field theory, it will tell you what you need to know about strings. Chapter 7 deals with string `dualities': some fascinating and powerful symmetries that arise when space has an interesting topology (a cylinder, say). We argue that they are the kind of symmetries are not merely observational, but `go all the way down', showing that string theory does not possess, in its basic objects, familiar spacetime properties, such as definite size or topology; it is for largely that reason that spacetime `emerges'. Chapter 8 is again fairly technical, explaining and analyzing in some detail the derivation of the Einstein field equation for gravity, from string theory. This is a central part of emergence, for it derives the spacetime metric, giving empirical content to spacetime geometry, and gives rise to GR. Finally, chapter 9 draws on the material of the previous chapters to argue that indeed spacetime emerges in string theory, how this happens, and what `principles of physical salience' are required.

The final, concluding, chapter draws on the results of the previous ones to return to the question of this introduction. How can we see that the derivations of spacetime that we have investigated are themselves physically salient, and what principles can we extract from them that might be helpful in the search for QG?

\bibliographystyle{plainnat}
\bibliography{../../Bibliography/biblio}

\begin{thebibliography}{49}
\providecommand{\natexlab}[1]{#1}
\providecommand{\url}[1]{\texttt{#1}}
\expandafter\ifx\csname urlstyle\endcsname\relax
  \providecommand{\doi}[1]{doi: #1}\else
  \providecommand{\doi}{doi: \begingroup \urlstyle{rm}\Url}\fi

\bibitem[Albert(2015)]{alb15}
David Albert.
\newblock \emph{After Physics}.
\newblock Harvard University Press, Cambridge, MA, 2015.

\bibitem[Albert(1996)]{Alb:96}
David~Z Albert.
\newblock Elementary quantum metaphysics.
\newblock In \emph{Bohmian mechanics and quantum theory: An appraisal}, pages
  277--284. Springer, 1996.

\bibitem[Barrett(1996)]{bar96}
Jeffrey~A Barrett.
\newblock Empirical adequancy and the availability of reliable records in
  quantum mechanics.
\newblock \emph{Philosophy of Science}, 63:\penalty0 49--64, 1996.

\bibitem[Bose et~al.(2017)Bose, Mazumdar, Morley, Ulbricht, Toro{\v{s}},
  Paternostro, Geraci, Barker, Kim, and Milburn]{BosMazMor:17}
Sougato Bose, Anupam Mazumdar, Gavin~W Morley, Hendrik Ulbricht, Marko
  Toro{\v{s}}, Mauro Paternostro, Andrew~A Geraci, Peter~F Barker, MS~Kim, and
  Gerard Milburn.
\newblock Spin entanglement witness for quantum gravity.
\newblock \emph{Physical Review Letters}, 119\penalty0 (24):\penalty0 240401,
  2017.

\bibitem[Busza et~al.(1999)Busza, Jaffe, Sandweiss, and
  Wilczek]{W.-Busza:1999uq}
W.~Busza, R.L. Jaffe, J.~Sandweiss, and F.~Wilczek.
\newblock Review of speculative ``disaster scenarios'' at rhic.
\newblock Technical report, Brookhaven National Laboratory,
  http://www.bnl.gov/rhic/docs/rhicreport.pdf, September 1999.

\bibitem[Butterfield(2011{\natexlab{a}})]{but11a}
Jeremy Butterfield.
\newblock Emergence, reduction and supervenience: A varied landscape.
\newblock \emph{Foundations of Physics}, 41:\penalty0 920--959,
  2011{\natexlab{a}}.

\bibitem[Butterfield(2011{\natexlab{b}})]{but11b}
Jeremy Butterfield.
\newblock Less is different: emergence and reduction reconciled.
\newblock \emph{Foundations of Physics}, 41:\penalty0 1065--1135,
  2011{\natexlab{b}}.

\bibitem[Butterfield and Gomes(2020{\natexlab{a}})]{ButGom:20}
Jeremy Butterfield and Henrique Gomes.
\newblock Functionalism as a species of reduction, July 2020{\natexlab{a}}.
\newblock URL \url{http://philsci-archive.pitt.edu/18043/}.
\newblock Submitted to: ?Current Debates in Philosophy of Science: In honor of
  Roberto Torretti?, edited by Cristian Soto; (to be published in the Synthese
  Library).

\bibitem[Butterfield and Gomes(2020{\natexlab{b}})]{GomBut:20}
Jeremy Butterfield and Henrique Gomes.
\newblock Geometrodynamics as functionalism about time, October
  2020{\natexlab{b}}.
\newblock URL \url{http://philsci-archive.pitt.edu/18339/}.
\newblock Submitted to: From Quantum to Classical: Essays in memory of Dieter
  Zeh; edited by Claus Kiefer: Springer, Cham, 2021.

\bibitem[Callender and Huggett(2001)]{Callender:2001uq}
Craig Callender and Nick Huggett.
\newblock Why quantize gravity (or any other field for that matter)?
\newblock \emph{Philosophy of Science}, 68:\penalty0 S382--94, 2001.

\bibitem[Carroll et~al.(2001)Carroll, Harvey, Kostelecky, Lane, and
  Okamoto]{carhar+:a}
Sean~M. Carroll, Jeffrey~A. Harvey, V.~Alan Kostelecky, Charles~D. Lane, and
  Takemi Okamoto.
\newblock Noncommutative field theory and lorentz violation.
\newblock 2001.
\newblock URL \url{http://arxiv.org/abs/hep-th/0105082v1}.

\bibitem[Chaichian et~al.(2000)Chaichian, Demichev, and
  Presnajder]{ChaDemPre:00}
M.~Chaichian, A.~Demichev, and P.~Presnajder.
\newblock Quantum field theory on noncommutative space-times and the
  persistence of ultraviolet divergences.
\newblock \emph{Nucl.Phys.}, B567:\penalty0 360--390, 2000.
\newblock URL \url{http://arxiv.org/abs/hep-th/9812180}.

\bibitem[Clarke et~al.(1956)Clarke, Leibniz, and Alexander]{LeiClaLei:56}
Samuel Clarke, Gottfried~Wilhelm Leibniz, and Robert~Gavin Alexander.
\newblock \emph{The Leibniz-Clarke Correspondence: Together Wiith Extracts from
  Newton's Principia and Opticks}.
\newblock Manchester University Press, 1956.

\bibitem[Crowther(2016)]{cro16}
Karen Crowther.
\newblock \emph{Effective Spacetime: Understanding Emergence in Effective Field
  Theory and Quantum Gravity}.
\newblock Springer, Cham, 2016.

\bibitem[Descartes(1644)]{Des:44}
Rene Descartes.
\newblock \emph{Principia Philosophiae}.
\newblock Apud Ludovicum Elezvirium, 1644.

\bibitem[DiSalle(2006)]{DiS:06}
Robert DiSalle.
\newblock \emph{Understanding Spacetime: The Philosophical Development of
  Physics from Newton to Einstein}.
\newblock Cambridge University Press, Cambridge, 2006.

\bibitem[Earman(1989)]{ear:89b}
John Earman.
\newblock \emph{World Enough and Space-Time: Absolute versus Relational
  Theories of Space and Time}.
\newblock MIT Press, Cambridge, MA, 1989.

\bibitem[Friedman(2001)]{Fri:01}
Michael Friedman.
\newblock \emph{Dynamics of Reason}.
\newblock CSLI Publications, 2001.

\bibitem[Gelfand and Naimark(1943)]{GelNai:43}
I.M Gelfand and M.A Naimark.
\newblock On the embedding of normed rings into the ring of operators in
  hilbert space.
\newblock \emph{Mat. Sbornik}, 12:\penalty0 197--213, 1943.

\bibitem[Geroch(1972)]{ger:72a}
Robert Geroch.
\newblock Einstein algebras.
\newblock \emph{Communications in Mathematical Physics}, 26:\penalty0 271--275,
  1972.

\bibitem[Hagar and Hemmo(2013)]{hagar2013primacy}
Amit Hagar and Meir Hemmo.
\newblock The primacy of geometry.
\newblock \emph{Studies in History and Philosophy of Science Part B: Studies in
  History and Philosophy of Modern Physics}, 44\penalty0 (3):\penalty0
  357--364, 2013.

\bibitem[Hawking(1974)]{Hawking:1974fk}
S.~W. Hawking.
\newblock Black hole explosions?
\newblock \emph{Nature}, 248\penalty0 (5443):\penalty0 30--31, 1974.

\bibitem[Hesse(1961)]{Hes:61}
Mary~B Hesse.
\newblock \emph{Forces and fields}.
\newblock T. Nelson, 1961.

\bibitem[Huggett(2012)]{Hug:12}
Nick Huggett.
\newblock What did newton mean by \textquoteleft{}absolute
  motion\textquoteright?
\newblock In Andrew Janiak and Eric Schliesser, editors, \emph{Interpreting
  Newton: Critical Essays}, pages 196--218. Cambridge University Press, 2012.

\bibitem[Huggett(2018)]{Hug:18}
Nick Huggett.
\newblock Spacetime 'emergence', December 2018.
\newblock URL \url{http://philsci-archive.pitt.edu/15440/}.
\newblock To appear in Routledge Companion to Philosophy of Physics, edited by
  Eleanor Knox and Alastair Wilson.

\bibitem[Huggett and W{\"u}thrich(2013)]{HugWut:13}
Nick Huggett and Christian W{\"u}thrich.
\newblock Emergent spacetime and empirical (in) coherence.
\newblock \emph{Studies in History and Philosophy of Science Part B: Studies in
  History and Philosophy of Modern Physics}, 44\penalty0 (3):\penalty0
  276--285, 2013.

\bibitem[Huggett et~al.(forthcoming)Huggett, Lizzi, and Menon]{HugLizMen:20}
Nick Huggett, Fedele Lizzi, and Tushar Menon.
\newblock Missing the point in noncommutative geometry.
\newblock \emph{Synthese}, forthcoming.

\bibitem[Kiefer(2004)]{Kie:12}
Claus Kiefer.
\newblock \emph{Quantum Gravity}.
\newblock Oxford University Press, 2004.

\bibitem[Kim(2005)]{kim05}
Jaegwon Kim.
\newblock \emph{Physicalism, or Something Near Enough}.
\newblock Princeton University Press, Princeton, 2005.

\bibitem[Knox(2013)]{knox13}
Eleanor Knox.
\newblock Effective spacetime geometry.
\newblock \emph{Studies in History and Philosophy of Modern Physics},
  44:\penalty0 346--356, 2013.

\bibitem[Knox(2014)]{knox14}
Eleanor Knox.
\newblock Spacetime structuralism or spacetime functionalism?
\newblock Manuscript, 2014.

\bibitem[Knox(2019)]{knox19}
Eleanor Knox.
\newblock Physical relativity from a functionalist perspective.
\newblock \emph{Studies in History and Philosophy of Modern Physics},
  67:\penalty0 118--124, 2019.

\bibitem[Kuhn(1962)]{Kuh:62}
Thomas~S Kuhn.
\newblock \emph{The Structure of Scientific Revolutions}.
\newblock University of Chicago Press, Chicago, 1962.

\bibitem[Lam and W\"uthrich(2018)]{lamwut18}
Vincent Lam and Christian W\"uthrich.
\newblock Spacetime is as spacetime does.
\newblock \emph{Studies in History and Philosophy of Modern Physics},
  64:\penalty0 39--51, 2018.

\bibitem[Lam and W\"uthrich(forthcoming)]{lamwut20}
Vincent Lam and Christian W\"uthrich.
\newblock Spacetime functionalism from a realist perspective.
\newblock \emph{Synthese}, forthcoming.

\bibitem[Le~Bihan(2018)]{blb18}
Baptiste Le~Bihan.
\newblock Priority monism beyond spacetime.
\newblock \emph{Metaphysica}, 19:\penalty0 95--111, 2018.

\bibitem[Leake(1999)]{Leake:1999fk}
Jonathan Leake.
\newblock Big bang machine could destroy earth.
\newblock \emph{The Sunday Times}, July 18 1999.

\bibitem[Lewis(1972)]{Lew:72}
David Lewis.
\newblock Psychophysical and theoretical identifications.
\newblock \emph{Australasian Journal of Philosophy}, 50\penalty0 (3):\penalty0
  249--258, 1972.

\bibitem[Lizzi(2009)]{Liz:09}
Fedele Lizzi.
\newblock Noncommutative spaces.
\newblock \emph{Lecture Notes in Physics}, 774:\penalty0 89--109, 2009.

\bibitem[Maudlin(2007)]{mau07}
Tim Maudlin.
\newblock Completeness, supervenience, and ontology.
\newblock \emph{Journal of Physics A: Mathematical and Theoretical},
  40:\penalty0 3151--3171, 2007.

\bibitem[Nastase(2005)]{Nas:05}
Horatiu Nastase.
\newblock The rhic fireball as a dual black hole.
\newblock \emph{arXiv preprint hep-th/0501068}, 2005.

\bibitem[Newton(1726)]{New:26}
Isaac Newton.
\newblock \emph{Philosophiae naturalis principia mathematica}, volume~3.
\newblock Apud Guil. \& Joh. Innys, Regi{\ae} Societatis typographos, 1726.

\bibitem[Newton(1730)]{New:30}
Isaac Newton.
\newblock \emph{Opticks}.
\newblock Prabhat Prakashan, 1730.

\bibitem[Ney(2015)]{ney15}
Alyssa Ney.
\newblock Fundamental physical ontologies and the constraint of empirical
  coherence.
\newblock \emph{Synthese}, 192:\penalty0 3105--3124, 2015.

\bibitem[Oriti(2014)]{Ori:14}
Daniele Oriti.
\newblock Disappearance and emergence of space and time in quantum gravity.
\newblock \emph{Studies in History and Philosophy of Science Part B: Studies in
  History and Philosophy of Modern Physics}, 46:\penalty0 186--199, 2014.

\bibitem[Sklar(1983)]{skl83}
Lawrence Sklar.
\newblock Prospects for a causal theory of space-time.
\newblock In Richard Swinburne, editor, \emph{Space, Time and Causality}, pages
  45--62. D.~Reidel Publishing Company, Dordrecht, 1983.

\bibitem[Wallace(2012)]{wal:12}
David Wallace.
\newblock \emph{The Emergent Multiverse: Quantum Theory According to the
  Everett Interpretation}.
\newblock Oxford University Press, Oxford, 2012.

\bibitem[Walsh and Button(2018)]{WalBut:18}
Sean Walsh and Tim Button.
\newblock \emph{Philosophy and Model Theory}.
\newblock Oxford, UK: Oxford University Press, 2018.

\bibitem[W\"uthrich(2005)]{wut05}
Christian W\"uthrich.
\newblock To quantize or not to quantize: fact and folklore in quantum gravity.
\newblock \emph{Philosophy of Science}, 72:\penalty0 777--788, 2005.

\end{thebibliography}

\end{document}